\documentclass{article}
\usepackage{graphicx}
\usepackage{amsmath}

\input{tcilatex}

\begin{document}

\title{CASIMIR EFFECT FOR A DIELECTRIC WEDGE}
\author{I. Brevik and K. Pettersen \\
Division of Applied Mechanics,\\
Norwegian University of Science and Technology,\\
N-7491 Trondheim, Norway}
\maketitle

\begin{abstract}
The Casimir effect is considered for a wedge with opening angle $\alpha $,
with perfectly conducting walls, when the interior region is filled with an
isotropic and nondispersive medium with permittivity $\epsilon $ and
permeability $\mu $. The electromagnetic energy-momentum tensor in the bulk
is calculated, together with the surface stress on the walls. A discussion
is given on the possibilities for measuring the influence of the medium, via
the Casimir-Polder force.
\end{abstract}


\pagebreak

\section{INTRODUCTION}

The wedge geometry is an attractive system to study in connection with the
Casimir effect, since the geometry is nontrivial enough to exhibit the
essentials of phenomenological quantum field theory in continuous media, and
yet so simple that it avoids the formal divergences that so often plague
specific calculations once curved boundaries are present. An additional
bonus from considering a system of this kind is that one experiences an
interesting formal analogy with the theory of a straight cosmic string.

In this paper we will consider the Casimir theory of a wedge-shaped region
of opening angle $\alpha $, when the walls located at angles $\theta =0$ and 
$\theta =\alpha $ are perfectly conducting, and the interior region $%
0<\theta <\alpha $ is filled with a homogeneous and isotropic dielectric
medium of constant permittivity $\epsilon $ and permeability $\mu $. Figure
1 shows the geometry. The cusp is at the origin.

The present paper is a generalization of earlier work [1], in which the
interior volume was taken to be a vacuum, and is also closely related to
Ref. [2]. As for references to the earlier literature, we mention those
listed in [1] and [2]. We may only recall here that our formalism is based
upon Schwinger's source theory, as developed for the strongly related case
of cylindrical geometry in [3] and [4], and is related to the general
formalism given in Stratton's book [5]. A somewhat different approach is
followed by Mostepanenko and Trunov in their book [6]; their Section 2.3
treats the specific wedge geometry in the vacuum case.

We ought to point out that the formulation of Casimir theory to include
material properties in the bulk does not imply merely an almost trivial
input of factors $\epsilon $ and $\mu $. The phenomenological
electrodynamics is generally quite different from electrodynamics in a
vacuum. In particular, the four-momentum of a photon in a medium as
constructed on basis of Minkowski's energy-momentum tensor - or equivalently
from the Hamiltonian approach - is spacelike, so it is possible to make the
electrodynamic field energy \textit{negative} by means of a Lorentz
transformation. (Physically, it is precisely properties of this sort that
are underlying the recent discussions on the so-called analog models of
general relativity; cf., for instance, the papers of Leonhardt and Piwnicki
[7] and the conference report in [8].)

In the following section we derive the expression for the fundamental dyad $%
\mathbf{\Gamma }$, from which the effective products of the fields can be
constructed. The two scalar Green functions, $F_{m}$ and $G_{m}$, are
determined. In Section 4, the electromagnetic energy-momentum tensor $\Theta
_{\mu \nu }$ is calculated. It is rather remarkable, as shown by Eq. (\ref
{37}), how the formalism conspires so as to \ give a very simple result: the
components of $\Theta _{\mu \nu }$ reflect the presence of the medium only
through a common prefactor $1/\sqrt{\epsilon \mu }$. In Section 5 we discuss
possibilities for measurements, in particular, how the deflection of an
atomic beam in a medium-filled wedge is influenced by the medium. Also, the
formal analogy with the theory of cosmic strings is briefly commented upon.

We put $\hbar =c=1$, and adopt electromagnetic Heaviside-Lorentz units.

\section{DYADIC SOLUTION}

Referring to the formalism developed in [1] and [2] for the vacuum fields in
the bulk, we can here be brief. We give only the basic definitions, and
write down formulas when they deviate from the vacuum case. The starting
point, as always when working with Schwinger's source theory, is the
relationship

\begin{equation}
\mathbf{E}\left( x\right) =\int dx^{\prime }\mathbf{\Gamma }\left(
x,x^{\prime }\right) \cdot \mathbf{P}\left( x\right)  \label{1}
\end{equation}
between the electric field $\mathbf{E}\left( x\right) $\ and the
polarization source $\mathbf{P}\left( x\right) $. Here $\mathbf{\Gamma }%
\left( x,x^{\prime }\right) $\ is the basic dyad in the formalism; its
Fourier transform $\mathbf{\Gamma }\left( \mathbf{r},\mathbf{r}^{\prime
},\omega \right) $ follows from 
\begin{equation}
\mathbf{\Gamma }\left( x,x^{\prime }\right) =\int_{-\infty }^{\infty }\frac{%
d\omega }{2\pi }e^{-i\omega \tau }\Gamma \left( \mathbf{r},\mathbf{r}%
^{\prime },\omega \right) ,  \label{2}
\end{equation}
with $\tau =t-t^{\prime }$. Maxwell's equations lead to the governing
equation 
\begin{equation}
\nabla \times \nabla \times \mathbf{\Gamma }\left( \mathbf{r},\mathbf{r}%
^{\prime },\omega \right) -\epsilon \mu \omega ^{2}\mathbf{\Gamma }\left( 
\mathbf{r},\mathbf{r}^{\prime },\omega \right) =-\mu \omega ^{2}\mathbf{1}%
\delta \left( \mathbf{r}-\mathbf{r}^{\prime }\right) ,  \label{3}
\end{equation}
where $\mathbf{1}$\ is the unit dyad. It is \ advantageous to introduce a
new dyad $\mathbf{\Gamma }^{\prime }$ which is divergence-free, 
\begin{equation}
\Gamma ^{\prime }=\Gamma +\frac{1}{\epsilon }\mathbf{1}\delta \left( \mathbf{%
r}-\mathbf{r}^{\prime }\right) ,\;\;\;\nabla \cdot \Gamma ^{\prime }=0.
\label{4}
\end{equation}
The effective electric and magnetic field products are (in addition to [1-4]
cf., for instance, also [9]): 
\begin{eqnarray}
i\left\langle E_{i}\left( \mathbf{r}\right) E_{j}\left( \mathbf{r}^{\prime
}\right) \right\rangle _{\omega } &=&\mathbf{\Gamma }_{ij}^{\prime }\left( 
\mathbf{r},\mathbf{r}^{\prime },\omega \right) ,  \label{5} \\
i\left\langle H_{i}\left( \mathbf{r}\right) H_{j}\left( \mathbf{r}^{\prime
}\right) \right\rangle _{\omega } &=&-\frac{1}{\mu ^{2}\omega ^{2}}\left(
\nabla \times \mathbf{\Gamma }^{\prime }\times \nabla ^{\prime }\right) _{ij}
\notag \\
&=&\frac{1}{\mu ^{2}\omega ^{2}}\varepsilon _{ikl}\varepsilon _{jmn}\partial
_{k}\partial _{m}^{\prime }\Gamma _{\text{ln}}^{\prime }\left( \mathbf{r},%
\mathbf{r}^{\prime },\omega \right) .  \label{6}
\end{eqnarray}
Angular brackets mean quantum mechanical expectation values. These effective
products are to be inserted into the electromagnetic energy-momentum tensor $%
\left\langle S_{\mu \nu }\right\rangle $. The spatial components of $S_{\mu
\nu }$\ are 
\begin{equation}
S_{ik}=-E_{i}D_{k}-H_{i}B_{k}+\frac{1}{2}\delta _{ik}\left( \mathbf{E\cdot D}%
+\mathbf{H\cdot B}\right) .  \label{7}
\end{equation}
As $\mathbf{\Gamma }^{\prime }$\ is divergence-free, it can conveniently be
expanded in cylindrical coordinates. This implies use of the vector
spherical harmonics, 
\begin{equation}
\mathbf{X}_{lm}\left( \Omega \right) =\left[ l\left( l+1\right) \right]
^{-1/2}\mathbf{L}Y_{lm}\left( \Omega \right)  \label{7a}
\end{equation}
As mentioned above we let $\theta $ denote the polar angle; the cusp of the
wedge coincides with the $z$ axis. The boundary conditions are that the
electric field is normal, and the magnetic field tangential, at $\theta
=0,\alpha $ (se Fig.1). We let radii $\mathbf{r}$\ and $\mathbf{r}^{\prime }$%
\ correspond to $\theta ,z$ and $\theta ^{\prime },z^{\prime }$\
respectively, and introduce new symbols $\nu _{m}$\ defined by 
\begin{equation}
\nu _{m}=\frac{m\pi }{\alpha },  \label{8}
\end{equation}
with $m$\ a non-negative integer. A lengthy calculation along the lines of
[1] and [2] leads to the following integral expressions for the spectral
dyad $\Gamma ^{\prime }$ and its double curl: 
\begin{eqnarray}
&&\mathbf{\Gamma }^{\prime }\left( \mathbf{r},\mathbf{r}^{\prime },\omega
\right) =\frac{2}{\alpha }\left. \sum_{m=0}^{\infty }\right. ^{\prime
}\int_{-\infty }^{\infty }\frac{dk}{2\pi }  \notag \\
&&\times \left[ -\frac{1}{\epsilon \omega ^{2}}\left( \nabla \times \widehat{%
z}\right) \left( \nabla ^{\prime }\times \widehat{z} \right) \left(
d_{m}-k^{2}\right) F_{m}\left( r,r^{\prime }\right) \cos \nu _{m}\theta \cos
\nu _{m}\theta ^{\prime }\right.  \notag \\
&&+\frac{1}{\epsilon \omega }\left( \nabla \times \nabla \times \widehat{z}%
\right) \left( \nabla ^{\prime }\times \nabla ^{\prime }\times \widehat{z}%
\right) G_{m}\left( r,r^{\prime }\right)  \notag \\
&&\times \left. \sin \nu _{m}\theta \sin \nu _{m}\theta ^{\prime }\right]
e^{ik\left( z-z^{\prime }\right) },  \label{9}
\end{eqnarray}
\begin{eqnarray}
&&\nabla \times \mathbf{\Gamma }^{\prime }\left( \mathbf{r},\mathbf{r}%
^{\prime },\omega \right) \times \nabla ^{\prime }=\frac{2}{\alpha }\left.
\sum_{m=0}^{\infty }\right. ^{\prime }\int_{-\infty }^{\infty }\frac{dk}{%
2\pi }  \notag \\
&&\times \left[ \frac{1}{\epsilon \omega ^{2}}\left( \nabla \times \nabla
\times \widehat{z}\right) \left( \nabla ^{\prime }\times \nabla ^{\prime
}\times \widehat{z} \right) \left( d_{m}-k^{2}\right) F_{m}\left(
r,r^{\prime }\right) \cos \nu _{m}\theta \cos \nu _{m}\theta ^{\prime
}\right.  \notag \\
&&-\frac{1}{\epsilon \omega }\left( \nabla \times \widehat{z}\right) \left(
\nabla ^{\prime }\times \widehat{z}\right) \left( d_{m}-k^{2}\right) \left(
d_{m}^{\prime }-k^{2}\right) G_{m}\left( r,r^{\prime }\right)  \notag \\
&&\times \left. \sin \nu _{m}\theta \sin \nu _{m}\theta ^{\prime }\right]
e^{ik\left( z-z^{\prime }\right) }.  \label{10}
\end{eqnarray}
Here $k\in $\ $\left\langle -\infty ,\infty \right\rangle $ is the axial
wave number, the prime on the summation sign means that the $m=0$ term is
taken with half weight, $\widehat{z}$\ is the unit vector in the $z$\
direction, and $F_{m}$, $G_{m}$\ are the two scalar Green functions.
Further, $d_{m}$\ is the differential operator 
\begin{equation}
d_{m}=\left( \frac{1}{r}\frac{\partial }{\partial r}r\frac{\partial }{%
\partial r}-\frac{\nu _{m}^{2}}{r^{2}}\right) .  \label{11}
\end{equation}
The scalar Green functions are explicitly 
\begin{eqnarray}
F_{m}\left( r,r^{\prime }\right) &=&\frac{\omega ^{2}}{q^{2}}\left[ \mathcal{%
G}_{m}^{F}\left( r,r^{\prime }\right) +\frac{i\pi }{2}J_{\nu _{m}}\left(
qr_{<}\right) H_{\nu _{m}}\left( qr_{>}\right) \right] ,  \label{12} \\
G_{m}\left( r,r^{\prime }\right) &=&\frac{\omega }{q^{2}}\left[ \mathcal{G}%
_{m}^{G}\left( r,r^{\prime }\right) +\frac{i\pi }{2}J_{\nu _{m}}\left(
qr_{<}\right) H_{\nu _{m}}\left( qr_{>}\right) \right] ,  \label{13}
\end{eqnarray}
where 
\begin{equation}
q^{2}=\epsilon \mu \omega ^{2}-k^{2},  \label{14}
\end{equation}
$J_{\nu _{m}}$ and $H_{\nu _{m}}$\ being ordinary Bessel and Hankel
functions of order $\nu _{m}$. Further, $\mathcal{G}_{m}^{F,G}$\ are the
so-called auxiliary Green functions, given by 
\begin{equation}
\mathcal{G}_{m}^{F,G}=-\frac{1}{2\nu _{m}}\left( \frac{r_{<}}{r_{>}}\right)
^{\nu _{m}}\;,\;\;m>0,  \label{15}
\end{equation}
\begin{equation}
\mathcal{G}_{0}^{F,G}=-\frac{1}{2}\ln \frac{r_{<}}{r_{>}}.  \label{16}
\end{equation}
It ought to be noted that the Green functions are derived on the basis of
requiring boundedness as $r\rightarrow 0$, and outgoing wave conditions as $%
r\rightarrow \infty $.

\section{EFFECTIVE FIELD PRODUCTS}

Using Eqs. (\ref{5}), (\ref{6}) and (\ref{9}), (\ref{10}) we can now
calculate the effective field product within the wedge, assuming that the
two spacetime points $x$ and $x^{\prime }$ are separated. In full
generality, all coordinates $\left\{ t,r,\theta ,z\right\} $ would be
different from $\left\{ t^{\prime },r^{\prime },\theta ^{\prime },z^{\prime
}\right\} $. However, we assume henceforth $t=t^{\prime }$\ and $z=z^{\prime
}$, so that $x$\ and $x^{\prime }$\ are separated only spatially, in the
radial and azimuthal directions.

It is first to be noted that the functions $\mathcal{G}_{0}^{F,G}$\ do not
contribute to the effective products. The argument runs similarly to that in
[1]. As for the auxiliary Green functions we are left only with $\mathcal{G}%
_{m}^{F,G}$\ with $m>0$.

Now introducing for convenience the operator 
\begin{equation}
\mathcal{L}=\frac{2}{\alpha }\frac{1}{\left( 2\pi \right) ^{2}}\int_{-\infty
}^{\infty }d\omega \int_{-\infty }^{\infty }dk\left. \sum_{m=0}^{\infty
}\right. ^{\prime },  \label{17}
\end{equation}
we obtain for the diagonal effective products 
\begin{eqnarray}
&&i\left\langle E_{\theta }\left( r,\theta \right) E_{\theta }\left(
r^{\prime },\theta ^{\prime }\right) \right\rangle =\mathcal{L}\left[ \frac{%
i\pi }{2}\left\{ \mu \omega ^{2}J_{\nu _{m}}^{\prime }\left( qr_{<}\right)
H_{\nu _{m}}^{\prime }\left( qr_{>}\right) \right. \right.  \notag \\
&&+\left. \frac{\nu _{m}^{2}k^{2}}{\epsilon q^{2}}\frac{1}{rr^{\prime }}%
J_{\nu _{m}}\left( qr_{<}\right) H_{\nu _{m}}\left( qr_{>}\right) \right\}
\left. \cos \nu _{m}\theta \cos \nu _{m}\theta ^{\prime }\right] ,
\label{18}
\end{eqnarray}
\begin{eqnarray}
&&i\left\langle H_{\theta }\left( r,\theta \right) H_{\theta }\left(
r^{\prime },\theta ^{\prime }\right) \right\rangle =\mathcal{L}\left[ \frac{%
i\pi }{2}\left\{ \epsilon \omega ^{2}J_{\nu _{m}}^{\prime }\left(
qr_{<}\right) H_{\nu _{m}}^{\prime }\left( qr_{>}\right) \right. \right. 
\notag \\
&&+\left. \left. \frac{\nu _{m}^{2}k^{2}}{\mu q^{2}}\frac{1}{rr^{\prime }}%
J_{\nu _{m}}\left( qr_{<}\right) H_{\nu _{m}}\left( qr_{>}\right) \right\}
\sin \nu _{m}\theta \sin \nu _{m}\theta ^{\prime }\right] ,  \label{19}
\end{eqnarray}
\begin{eqnarray}
&&i\left\langle E_{r}\left( r,\theta \right) E_{r}\left( r^{\prime },\theta
^{\prime }\right) \right\rangle =\mathcal{L}\left[ \frac{i\pi }{2}\left\{ 
\frac{k^{2}}{\epsilon }J_{\nu _{m}}^{\prime }\left( qr_{<}\right) H_{\nu
_{m}}^{\prime }\left( qr_{>}\right) \right. \right.  \notag \\
&&+\left. \left. \frac{\nu _{m}^{2}\mu \omega ^{2}}{q^{2}}\frac{1}{%
rr^{\prime }}J_{\nu _{m}}\left( qr_{<}\right) H_{\nu _{m}}\left(
qr_{>}\right) \right\} \sin \nu _{m}\theta \sin \nu _{m}\theta ^{\prime }%
\right] ,  \label{20}
\end{eqnarray}
\begin{eqnarray}
&&i\left\langle H_{r}\left( r,\theta \right) H_{r}\left( r^{\prime },\theta
^{\prime }\right) \right\rangle =\mathcal{L}\left[ \frac{i\pi }{2}\left\{ 
\frac{k^{2}}{\mu }J_{\nu _{m}}^{\prime }\left( qr_{<}\right) H_{\nu
_{m}}^{\prime }\left( qr_{>}\right) \right. \right.  \notag \\
&&+\left. \left. \frac{\nu _{m}^{2}\epsilon \omega ^{2}}{q^{2}}\frac{1}{%
rr^{\prime }}J_{\nu _{m}}\left( qr_{<}\right) H_{\nu _{m}}\left(
qr_{>}\right) \right\} \cos \nu _{m}\theta \cos \nu _{m}\theta ^{\prime }%
\right] ,  \label{21}
\end{eqnarray}
\begin{equation}
i\left\langle E_{z}\left( r,\theta \right) E_{z}\left( r^{\prime },\theta
^{\prime }\right) \right\rangle =\mathcal{L}\left[ \frac{i\pi }{2}\frac{q^{2}%
}{\epsilon }J_{\nu _{m}}\left( qr_{<}\right) H_{\nu _{m}}\left(
qr_{>}\right) \sin \nu _{m}\theta \sin \nu _{m}\theta ^{\prime }\right] ,
\label{22}
\end{equation}
\begin{eqnarray}
&&i\left\langle H_{z}\left( r,\theta \right) H_{z}\left( r^{\prime },\theta
^{\prime }\right) \right\rangle \\
&=&\mathcal{L}\left[ \frac{i\pi }{2}\frac{q^{2}}{\mu }J_{\nu _{m}}\left(
qr_{<}\right) H_{\nu _{m}}\left( qr_{>}\right) \cos \nu _{m}\theta \cos \nu
_{m}\theta ^{\prime }\right] .
\end{eqnarray}
These are the diagonal products needed to calculate the normal stresses on
the surface. There are nondiagonal field products also, $\left\langle
E_{i}E_{k}\right\rangle $\ and $\left\langle H_{i}H_{k}\right\rangle $, with 
$i\neq k$. These products do not vanish by themselves, but it is notable
that the sum $\epsilon \left\langle E_{i}E_{k}\right\rangle +\mu
\left\langle H_{i}H_{k}\right\rangle $, present in the electromagnetic
stress tensor, does vanish.

It is also to be noted that the auxiliary Green functions $\mathcal{G}%
_{m}^{F,G}$\ with $m>0$\ do not contribute to the field products. This is
because of the relations $d_{m}\mathcal{G}_{m}^{F,G}=0$, as well as 
\begin{equation}
\left( d_{m}+\epsilon \mu \omega ^{2}-k^{2}\right) F_{m}=\omega ^{2}\mathcal{%
G}_{m}^{F},  \label{24}
\end{equation}
\begin{equation}
\left( d_{m}+\epsilon \mu \omega ^{2}-k^{2}\right) G_{m}=\omega \mathcal{G}%
_{m}^{G}.  \label{25}
\end{equation}
Finally, one may verify by explicit calculation that the Poynting vector
vanishes. This is as one would expect, under stationary conditions.

It is to be emphasized that the differences $\left( r-r^{\prime }\right) $\
and $\left( \theta -\theta ^{\prime }\right) $\ are at the present stage of
the calculation arbitrary; they are not necessarily small.

\section{ENERGY-MOMENTUM TENSOR}

We now let the two points $r$\ and $r^{\prime }$\ approach each other, but
keep both $\left( r-r^{\prime }\right) $ and $\left( \theta -\theta ^{\prime
}\right) $\ different from zero. For convenience we write henceforth $%
\left\langle E_{r}^{2}\right\rangle $\ instead of $\left\langle E_{r}\left(
r,\theta \right) E_{r}\left( r^{\prime },\theta ^{\prime }\right)
\right\rangle _{\mathbf{r}\rightarrow \mathbf{r}^{\prime }}$, etc. The
azimuthal diagonal component of $\left\langle S_{\mu \nu }\right\rangle $\
at an arbitrary position $\mathbf{r}$\ within the wedge is 
\begin{eqnarray}
\left\langle S_{\theta \theta }\left( \mathbf{r}\right) \right\rangle &=&%
\frac{1}{2}\left[ \epsilon \left\langle E_{r}^{2}\right\rangle -\epsilon
\left\langle E_{\theta }^{2}\right\rangle +\epsilon \left\langle
E_{z}^{2}\right\rangle \right.  \notag \\
&&+\left. \mu \left\langle H_{r}^{2}\right\rangle -\mu \left\langle
H_{\theta }^{2}\right\rangle +\mu \left\langle H_{z}^{2}\right\rangle \right]
.  \label{26}
\end{eqnarray}
Inserting the effective products above we get 
\begin{eqnarray}
\left\langle S_{\theta \theta }\left( \mathbf{r}\right) \right\rangle &=&%
\frac{\pi }{4}\mathcal{L}\left[ \left\{ q^{2}+\frac{1}{rr^{\prime }}\frac{%
\partial }{\partial \theta }\frac{\partial }{\partial \theta ^{\prime }}-%
\frac{\partial }{\partial r}\frac{\partial }{\partial r^{\prime }}\right\}
\right.  \notag \\
&&\times \left. J_{\nu _{m}}\left( qr_{<}\right) H_{\nu _{m}}\left(
qr_{>}\right) \cos \nu _{m}\left( \theta -\theta ^{\prime }\right) \right] _{%
\mathbf{r}\rightarrow \mathbf{r}^{\prime }}.  \label{27}
\end{eqnarray}
From this we have to subtract off the contact term, called $\left\langle
S_{\theta \theta }^{0}\left( \mathbf{r}\right) \right\rangle $. As the
Casimir effect is caused by the boundaries of the wedge, it follows that the
contact term has to be evaluated in the absence of any boundaries at all. In
other words, $\left\langle S_{\theta \theta }^{0}\left( \mathbf{r}\right)
\right\rangle $\ corresponds to a homogeneous dielectric extending over all
space. Explicit calculation shows that the contact term becomes equal to the
wedge effective product evaluated at $\alpha =\pi $: 
\begin{equation}
\left\langle S_{\theta \theta }^{0}\left( \mathbf{r}\right) \right\rangle
=\left\langle S_{\theta \theta }^{\alpha =\pi }\left( \mathbf{r}\right)
\right\rangle .  \label{28}
\end{equation}
Altough this is a natural result, it could hardly have been written down
beforehand, without explicit calculation.

We perform a complex frequency rotation, $\omega \rightarrow i\widehat{%
\omega }$, implying 
\begin{equation}
q=\sqrt{\epsilon \mu {\omega }^{2}-k^{2}}\rightarrow \sqrt{-\left(
\epsilon \mu \widehat{\omega }^{2}+k^{2}\right) }\equiv i\rho .  \label{29}
\end{equation}
Then, 
\begin{eqnarray}
&&\left\langle S_{\theta \theta }\left( \mathbf{r}\right) \right\rangle =%
\frac{1}{\alpha \pi ^{2}}\int_{-\infty }^{\infty }d\widehat{\omega }%
\int_{0}^{\infty }dk\left. \sum_{m=0}^{\infty }\right. ^{\prime }\left[
\left\{ -\rho ^{2}+\frac{1}{rr^{\prime }}\frac{\partial }{\partial \theta }%
\frac{\partial }{\partial \theta ^{\prime }}-\frac{\partial }{\partial r}%
\frac{\partial }{\partial r^{\prime }}\right\} \right.  \notag \\
&&\times \left. I_{\nu _{m}}\left( \rho r_{<}\right) K_{\nu _{m}}\left( \rho
r_{>}\right) \cos \nu _{m}\left( \theta -\theta ^{\prime }\right) \right] _{%
\mathbf{r}\rightarrow \mathbf{r}^{\prime }},  \label{30}
\end{eqnarray}
where $I_{\nu _{m}}$ and $K_{\nu _{m}}$\ are modified Bessel functions. We
go over to polar coordinates, noting that $\sqrt{\epsilon \mu }d\widehat{%
\omega }dk\rightarrow \rho d\rho d\phi $ since $\rho $\ is the radius in the 
$\sqrt{\epsilon \mu }\widehat{\omega },k$\ plane. Integrating over angles $%
\phi $\ from $0$\ to $\pi /2$\ we obtain 
\begin{eqnarray}
&&\left\langle S_{\theta \theta }\left( \mathbf{r}\right) \right\rangle = 
\notag \\
&&\frac{1}{2\pi \alpha \sqrt{\epsilon \mu }}\left. \sum_{m=0}^{\infty
}\right. ^{\prime }\int_{0}^{\infty }\rho d\rho \left[ \left\{ -\rho ^{2}+%
\frac{1}{rr^{\prime }}\frac{\partial }{\partial \theta }\frac{\partial }{%
\partial \theta ^{\prime }}-\frac{\partial }{\partial r}\frac{\partial }{%
\partial r^{\prime }}\right\} \right.  \notag \\
&&\times \left. I_{\nu _{m}}\left( \rho r_{<}\right) K_{\nu _{m}}\left( \rho
r_{>}\right) \cos \nu _{m}\left( \theta -\theta ^{\prime }\right) \right] _{%
\mathbf{r}\rightarrow \mathbf{r}^{\prime }}.  \label{31}
\end{eqnarray}
We now introduce 
\begin{equation}
p=\frac{\pi }{\alpha },  \label{32}
\end{equation}
and assume henceforth that $p$\ is an integer. Therewith $\nu _{m}=mp$ also
becomes an integer. This simplifying case is convenient, as it allows us to
make use of the generalized Graf addition theorem for modified Bessel
functions. The important formula in our context is given by Eq. (B.6) in
[1], and will not be repeated here. By means of it, we get 
\begin{eqnarray}
&&\left\langle S_{\theta \theta }\left( \mathbf{r}\right) \right\rangle =%
\frac{1}{\left( 2\pi \right) ^{2}\sqrt{\epsilon \mu }}\int_{0}^{\infty }\rho
d\rho  \notag \\
&&\times \left. \left[ -\rho ^{2}+\frac{1}{rr^{\prime }}\frac{\partial }{%
\partial \theta }\frac{\partial }{\partial \theta ^{\prime }}-\frac{\partial 
}{\partial r}\frac{\partial }{\partial r^{\prime }}\right]
\sum_{m=0}^{p-1}K_{0}\left( \rho R_{n}\right) \right| _{\mathbf{r}%
\rightarrow \mathbf{r}^{\prime }},  \label{33}
\end{eqnarray}
where 
\begin{equation}
R_{n}=\left[ r^{2}+r^{\prime 2}-2rr^{\prime }\cos \left( \left( \theta
-\theta ^{\prime }\right) +\frac{2\pi n}{p}\right) \right] ^{1/2}.
\label{34}
\end{equation}
This expression can be processed further, using Eqs. (56) - (60) in [1]. The
regularized energy-momentum tensor $\left\langle \Theta _{\mu \nu }\left( 
\mathbf{r}\right) \right\rangle $, as defined generally by 
\begin{equation}
\left\langle \Theta _{\mu \nu }\left( \mathbf{r}\right) \right\rangle
=\left\langle S_{\mu \nu }\left( \mathbf{r}\right) \right\rangle
-\left\langle S_{\mu \nu }^{0}\left( \mathbf{r}\right) \right\rangle ,
\label{35}
\end{equation}
then yields for the $\theta \theta $ component 
\begin{equation}
\left\langle \Theta _{\theta \theta }\left( \mathbf{r}\right) \right\rangle
=-\frac{3}{720\pi ^{2}\sqrt{\epsilon \mu }r^{4}}\left( \frac{\pi ^{2}}{%
\alpha ^{2}}+11\right) \left( \frac{\pi ^{2}}{\alpha ^{2}}-1\right) .
\label{36}
\end{equation}
Similar considerations can be carried out for the other components of the
energy-momentum tensor. If we numerate the components according to $%
\left\langle \Theta _{\mu \nu }\right\rangle =$\ $\left\langle \Theta
_{rr},\Theta _{\theta \theta },\Theta _{zz},-w\right\rangle $ where $w$\ is
the electromagnetic energy density, we get finally 
\begin{equation}
\left\langle \Theta _{\mu \nu }\left( \mathbf{r}\right) \right\rangle =\frac{%
1}{720\pi ^{2}\sqrt{\epsilon \mu }r^{4}}\left( \frac{\pi ^{2}}{\alpha ^{2}}%
+11\right) \left( \frac{\pi ^{2}}{\alpha ^{2}}-1\right) diag\left(
1,-3,1,1\right) .  \label{37}
\end{equation}
These expressions all vanish for $\pi =\alpha $, as expected.

The simplicity of the expression (\ref{37}) is rather remarkable. The
expression differs from the corresponding expression in vacuum [1] only
through the factor $\sqrt{\epsilon \mu }$ in the denominator. A value $n=%
\sqrt{\epsilon \mu }>1$ of the refractive index $n$ thus causes the energy
density in the bulk, as well as the normal stress on the plates, to be less
than the vacuum value. In the special case of\ $\epsilon \mu =1$, a case
considered repeatedly in recent years in various contexts (a so-called
''relativistic'' medium), one ends up with precisely the same
energy-momentum tensor as in vacuum. There seems to be no simple way to see
beforehand why the dependence on the properties of the medium should have
the special form of Eq. (\ref{37}).

\section{ON POSSIBILITIES FOR EXPERIMENTS}

We consider first the normal surface density on the lower wall, $\theta =0$,
as a function of the distance $r$ from the cusp. For convenience we will
denote this force density by $\sigma \left( r\right) $. Since our triplet of
basis vectors in the spatial directions $\left\{ r,\theta ,z\right\} $\ is
orthonormal, $\sigma \left( r\right) $\ must simply be equal to $%
-\left\langle S_{\theta \theta }\left( \mathbf{r}\right) \right\rangle $,
taken at the wall. Thus, in dimensional units, 
\begin{equation}
\sigma \left( r\right) =\frac{\hbar c}{720\pi ^{2}\sqrt{\epsilon \mu }r^{4}}%
\left( \frac{\pi ^{2}}{\alpha ^{2}}+11\right) \left( \frac{\pi ^{2}}{\alpha
^{2}}-1\right) .  \label{38}
\end{equation}
The force between the walls is attractive, as expected, and it decreases
quickly when one moves away from the cusp. The divergence at the cusp is
clearly fictitious: the presence of a skin depth $\delta $\ in the material
forbids us to apply the continuum dielectric model at very small distances.
Assume copper, for instance, for which the conductivity is $6.0\times
10^{7}\left( \Omega m\right) ^{-1}$. As the most significant frequencies
contributing to the Casimir force are of the order $c/a$, where $a$ is the
local separation between the walls, we obtain $\delta \sim 10\,nm$ if $a\sim
1\,\mu m$ (the typical separation distance) [1]. The expression (\ref{38})
is hardly applicable until $r$ becomes as large as about $1\,\mu m$.

Assume for definiteness a very narrow wedge, corresponding to $\alpha
=10^{-4}$rad (0.0057$^\circ$). A local wall separation of $a=1\,\mu m$ then
corresponds to $r=a/\alpha =1\,cm$. From Eq. (\ref{38}) we get 
\begin{equation}
\sigma \left( r=1\,cm\right) =\frac{0.0043}{\sqrt{\epsilon \mu }}\frac{dyn}{%
cm^{2}}
\end{equation}
This is about $\left( 3\sqrt{\epsilon \mu }\right) ^{-1}$ of the
conventional surface force density $0.013\,dyn/cm^{2}$ between two parallel
perfectly conducting plates at the same separation as above, $a=1\,\mu m$.

It does not seem to be easy to design a surface force experiment of this
kind. A more promising possibility might be to consider a variant of the
deflection experiment of Sukenik \textit{et al.} [10]. These authors
measured the deflection of a ground-state atomic beam passing through a
wedge-shaped cavity of opening angle $\alpha \sim 10^{-4}rad$, thus of the
same order of magnitude as considered above. Let us assume the wedge region
to be completely filled with a fluid, having material constants $\epsilon $\
and $\mu $. The transverse force on the beam, causing the deflection, will
clearly be $\epsilon $\ and $\mu $ dependent. This transverse force is of
course the Casimir-Polder rather than the Casimir force in the strict sense,
but the two kinds of forces are strongly interrelated.

Let us calculate the interaction energy $U\left( \mathbf{r}\right) $ for a
dipolar particle in the wedge as if the particle were at rest, at position $%
\mathbf{r}$. It is natural to assume that retardation effects are not
important, so that we can use the material properties of the medium at $%
\omega =0$ with satisfactory accuracy. Thus, we can use the \textit{static}
polarizability $\alpha \left( 0\right) $ for the particle, and so obtain 
\begin{equation}
U\left( \mathbf{r}\right) =-\frac{1}{2}\alpha \left( 0\right) \left\langle 
\mathbf{E}^{2}\right\rangle =-\frac{1}{2}\alpha \left( 0\right) \left[
\left\langle E_{r}^{2}\right\rangle +\left\langle E_{\theta
}^{2}\right\rangle +\left\langle E_{z}^{2}\right\rangle \right] .  \label{40}
\end{equation}
Using Eqs. (\ref{18}), (\ref{20}) and (\ref{22}) we obtain after some
calculation, putting $\theta =\theta ^{\prime }$, 
\begin{eqnarray}
U\left( \mathbf{r}\right) &=&-\frac{\alpha \left( 0\right) }{4\pi \alpha 
\sqrt{\epsilon \mu }}\frac{1}{\epsilon }\int_{0}^{\infty }\rho ^{3}d\rho 
\notag \\
&&\times \left. \sum_{m=0}^{\infty }\right. ^{\prime }\left[ I_{mp}^{\prime
}\left( mp\theta \right) K_{mp}^{\prime }\left( mp\theta \right) \cos
2mp\theta \right.  \notag \\
&&-\frac{m^{2}p^{2}}{\rho ^{2}rr^{\prime }}I_{mp}\left( \rho r_{<}\right)
K_{mp}\left( \rho r_{>}\right) \cos 2mp\theta  \notag \\
&&-\left. 2I_{mp}\left( \rho r_{<}\right) K_{mp}\left( \rho r_{>}\right)
\sin ^{2}mp\theta \right] .  \label{41}
\end{eqnarray}
Recall that $p=\pi /\alpha $ has been assumed to be an integer. On physical
grounds, $r$ and $r^{\prime }$ have to lie close to each other in the
expression (\ref{41}). The expression may be regularized: if one subtracts
off a contact term corresponding to $p=1$ ($\alpha =\pi $), one is left with
the wedge-specific contribution, equal to zero in the case of a single plane
plate in interaction with the dipole.

The expression (\ref{41}) has the important property that the influence from
the medium turns up only in the prefactor $\left( \sqrt{\epsilon \mu }%
\epsilon \right) ^{-1}$. That is, we can make use of the results derived
earlier for the case \ of a vacuum wedge [2]. Defining $\xi =r_{<}/r_{>}$
and taking $\left( \xi -1\right) $ to be small, we obtain using Eq. (3.10)
in [2]: 
\begin{eqnarray}
U\left( \mathbf{r}\right) &=&-\frac{\alpha \left( 0\right) }{16\pi ^{2}\sqrt{%
\epsilon \mu }\epsilon r^{4}}  \notag \\
&&\times \left[ \frac{3}{2}\frac{p^{4}}{\sin ^{4}p\theta }-\frac{p^{2}\left(
p^{2}-1\right) }{\sin ^{2}p\theta }-\frac{1}{90}\left( p^{2}+11\right)
\left( p^{2}-1\right) \right] .  \label{42}
\end{eqnarray}
Assuming $\sqrt{\epsilon \mu }\epsilon >1$ which is usually the case, we
thus find a weaker deflection of the point dipole toward the nearby wall
than in the case of a vacuum cavity. In the special case of a plane plate ($%
p=1$), only the first term in (\ref{42}) survives.

\section{CONCLUSIONS AND FURTHER REMARKS}

Let us first summarize:

\bigskip

\textbf{(1)} We assumed nondispersive permittivity $\epsilon $\ and
permeability $\mu $\ in the medium-filled wedge. All calculations were made
at zero temperature. The electromagnetic energy-momentum tensor $%
\left\langle \Theta _{\mu \nu }\right\rangle $, after regularization, is
given by Eq. (\ref{37}). All components of $\left\langle \Theta _{\mu \nu
}\right\rangle $\ are independent of the polar angle $\theta $\ and diminish
with distance $r$\ from the cusp as $r^{-4}$. The influence from the medium
turns up solely in the prefactor $\sqrt{\epsilon \mu }$ in the denominator.

\bigskip

\textbf{(2)} There exists to our knowledge no experiment testing the present
kind of theory. The most promising possibility seems to be the dielectric
variant of the experiment of Sukenik \textit{et al.} [10], measuring the
deflection of an atomic beam passing through a wedge-shaped cavity of small
opening angle $\alpha $. If the wedge is filled with a medium, the
transverse force on a dipole is given by the gradient of the potential (\ref
{42}), showing that the presence of the medium turns up solely in the
prefactor $\sqrt{\epsilon \mu }\epsilon $ in the denominator.

\bigskip

Then a couple of further remarks:

\bigskip

\textbf{(3)} The formulation of the theory at finite temperatures can be
carried out in the conventional way, replacing the integral over imaginary
frequencies $\widehat{\omega }$ by a sum over discrete Matsubara frequencies 
$\widehat{\omega }_{n}=\left( 2\pi nk_{B}T\right) $, with $n$ an integer.
This procedure was shown in [1] in detail for the component $\left\langle
S_{\theta \theta }\left( \mathbf{r}\right) \right\rangle ^{T}$, thus
generalizing the expression (\ref{30}) above to the case of finite
temperatures, and will not be further considered here. Again, the important
point in our context is the apperance of the extra factor $\sqrt{\epsilon
\mu }$ in the denominator in the energy-momentum tensor expression.

\bigskip

\textbf{(4)} Finally, it is worth noticing that the interesting formal
analogy that exists between a wedge and a straight cosmic string can be
carried over to the medium case, only with a slight modification. The line
element outside a string is 
\begin{equation}
ds^{2}=-dt^{2}+dr^{2}+\left( 1-4G\mu \right) ^{2}r^{2}d\theta ^{2}+dz^{2},
\label{43}
\end{equation}
where $G$\ is the gravitational constant and $\mu \simeq 10^{22}g/cm$ is the
string mass per unit length (GUT scale). Defining the symbol $\beta =\left(
1-4G\mu \right) ^{-1}$, we can write the electromagnetic energy-momentum
tensor as [11] 
\begin{equation}
\left\langle \Theta _{\mu \nu }\right\rangle =\frac{1}{720\pi ^{2}r^{4}}%
\left( \beta ^{2}+11\right) \left( \beta ^{2}-1\right) diag\left(
1,-3,1,1\right) .  \label{44}
\end{equation}
Comparison with (\ref{37}) shows that, apart from the extra prefactor $\sqrt{%
\epsilon \mu }$\ in the denominator of that equation, there is complete
analogy between the two cases if the gravitational quantity $\beta $\ is
identified with the wedge quantity $\pi /\alpha $.

\pagebreak


\textbf{REFERENCES}

[1] \ I. Brevik and M. Lygren, \textit{Ann. Phys. (N.Y.)} \textbf{251}
(1996), 157.

[2] \ I. Brevik, M. Lygren and V. N. Marachevsky,

\textit{Ann. Phys. (N.Y.)} \textbf{267} (1998), 134.

[3] \ I. Brevik and G. H. Nyland,

\textit{Ann. Phys. (N.Y.)} \textbf{230} (1994), 321.

[4] \ L. L. DeRaad, Jr. and K. A. Milton,

\textit{Ann. Phys. (N.Y.)} \textbf{136} (1981), 229.

[5] \ J. A. Stratton,

''Electromagnetic Theory'', McGraw - Hill, New York, 1941.

[6] \ V. M. Mostepanenko and N. N. Trunov,

''The Casimir Effect and its Applications'', Clarendon Press, Oxford, 1997.

[7] \ U. Leonhardt and P. Piwnicki,

\textit{Phys. Rev. A} \textbf{60} (1999), 4301; \textit{Phys. Rev. Lett.} 
\textbf{84} (2000), 822.

[8] \ Workshop on ''Analog models of General Relativity'',

Rio de Janeiro, Brazil, October 2000. Address: www.lafex.cbpf.br/\symbol{126}%
bscg/analog/

[9] \ K. A Milton,

\textit{Ann. Phys. (N.Y.)} \textbf{127} (1980), 49.

[10] \ C. I. Sukenik, M. G. Boshier, D. Cho, V. Sandoghdar and E. A. Hinds,

\textit{Phys Rev. Lett.} \textbf{70} (1993), 560.

[11] \ V. P. Frolov and E. M. Serebriany,

\textit{Phys. Rev. D} \textbf{35} (1987), 3779.

\end{document}